\documentclass[aps,prb,twocolumn,10pt]{revtex4-2}
\usepackage[utf8]{inputenc}
\usepackage{libertine}
\usepackage[T1]{fontenc}
\usepackage[libertine,upint]{newtxmath} 
\usepackage{graphicx,microtype,mathtools,bm,siunitx,booktabs,xcolor}
\usepackage[cal=boondoxo,frak=boondox]{mathalfa}
\definecolor{linkscolor}{cmyk}{0.6, 0.3, 0, 0.9}
\usepackage[colorlinks=true,allcolors=linkscolor!60]{hyperref}
\DeclareSIUnit\angstrom{\text{Å}}

\makeatletter
\g@addto@macro\@floatboxreset\centering
\makeatother 

\begin{document}

\title{Spin-dependent electron-electron interaction in Rashba materials}

\author{Yasha Gindikin}

\affiliation{Department of Condensed Matter Physics, Weizmann Institute of Science, Rehovot, 76100, Israel}

\author{Vladimir A.\ Sablikov}

\affiliation{Kotelnikov Institute of Radio Engineering and Electronics, Russian Academy of Sciences,\\ Fryazino branch, Fryazino, 141190, Russia}

\begin{abstract}
We review the effects of the pair spin-orbit interaction (PSOI) in Rashba materials. The PSOI is the electron-electron interaction component that depends on the spin and momentum of the electrons. Being produced by the Coulomb fields of interacting electrons, it exists already in vacuum, but becomes orders of magnitude larger in materials with the giant Rashba effect. The main nontrivial feature of the PSOI is that it is attractive for electrons in certain spin configurations tied to their momentum and competes with the Coulomb repulsion of the electrons. Under certain conditions attainable in modern low-dimensional structures the PSOI prevails. The resulting attraction between electrons leads to the formation of bound electron pairs, the binding energy of which can be controlled by electrical means. In many-electron systems the PSOI results in the instabilities of the uniform ground state with respect to the density fluctuations, which develop on different spatial scales, depending on the geometry of the electric fields that produce the PSOI\@. If the PSOI is not too strong the electronic system is stable, but its collective excitations reveal the highly unusual spin-charge structure and spectrum, which manifest themselves in the frequency dependence of the dynamic conductivity.
\end{abstract}

\maketitle

\section{Introduction}
From the early days of relativistic quantum mechanics it is known that the electrostatic Coulomb potential is insufficient to describe the interaction between electrons, and in the first quasi-relativistic approximation the potential of the electron-electron (e-e) interaction in addition to the purely Coulomb component $\mathcal{U}(\bm{r})$ contains a contribution that depends on the spin of the electrons and on their momenta~\cite{bethe2012quantum},
\begin{equation}
\label{hampsoi}
    \hat{H}_{\mathrm{PSOI}} = \frac{\alpha}{\hbar} \sum_{i \ne j} \left( \bm{\mathcal{E}} (\bm{r}_{i} - \bm{r}_{j}) \times \hat{\bm{p}}_{i}  \right) \cdot \hat{\bm{\sigma}}_{i}\,,
\end{equation}
where $\bm{\mathcal{E}}(\bm{r}) = \frac{1}{e} \nabla_{\!\bm{r}} \mathcal{U}(\bm{r})$ is the Coulomb field of e-e interaction, $\hat{\bm{p}}_i$ is the momentum operator of the $i$-th electron, $\hat{\bm{\sigma}} \equiv (\sigma_x,\sigma_y,\sigma_z)$ is the Pauli vector. This contribution is usually referred to as the pair spin-orbit interaction (PSOI). In vacuum its magnitude $\alpha = e \lambdabar^2/4$ is set by the square of the Compton length $\lambdabar$, that is relativistically small.

In crystalline solids the situation fundamentally changes due to the features of the band states, which under certain conditions lead to an extremely strong spin-orbit interaction (SOI) that depends on the electric field external to the crystalline one, including the Coulomb field of interacting electrons. This typically occurs as a result of the combined effect of momentum-dependent mixing of electron and hole subbands split by intra-atomic SOI and symmetry breaking created by the electric field~\cite{bychkov1984properties}. At present, the circle of these Rashba materials is very wide and the SOI parameter reaches gigantic values~\cite{manchon2015new,bihlmayer2015focus}.

The Rashba constant $\alpha$ of the currently known materials with giant SOI varies from $\SI{e2}{e \angstrom^2}$ in $\mathrm{InAs}$ to $\SI{e3}{e \angstrom^2}$  in such materials as $\mathrm{Bi}_2 \mathrm{Se}_3$~\cite{PhysRevLett.107.096802}, the monolayers of $\mathrm{BiSb}$~\cite{PhysRevB.95.165444}, the oxide heterostructures and films~\cite{varignon2018new}. In such materials, mainly two-dimensional ones, the PSOI is described by the same Hamiltonian as of Eq.~\eqref{hampsoi}, and for a sufficiently smooth potential $\mathcal{U}(r)$ the value of the $\alpha$ parameter is estimated at the level of the Rashba constant in a given material.

A distinctive feature of the PSOI as compared to the Coulomb interaction is that it is determined by the electric field ${\mathcal{E}}({r}) \sim r^{-2}$, which grows faster than the potential $\mathcal{U}(\bm{r}) \sim r^{-1}$ when we bring two electrons closer together. Hence there appears a new characteristic scale of $r=\sqrt{\alpha/e}$ below which the PSOI prevails over the Coulomb interaction. This scale can be large enough in modern heterostructures based on $\mathrm{LaAlO_3/SrTiO_3}$~\cite{Levy_2015,doi:10.1021/acs.nanolett.8b01614}, in specifically crafted structures based on two-dimensional (2D) layers of van der Waals materials with heavy adatoms~\cite{otrokov2018evidence,PhysRevB.99.085411}, etc. 

Most importantly, several related scales due to the competition of the PSOI and Coulomb interaction arise in many-electron systems, regulated by the electron concentration, density of states, and a particular geometry of the electric fields that generate the PSOI\@. The dependence of the pair interaction on the geometric configuration of the electric fields is not uncommon, for example, in the physics of excitons in thin films~\cite{rytova,keldysh1979coulomb}, but here it is much more pronounced as the PSOI is generated not by a scalar potential, but by a vector field, which gives rise to more possibilities.

Qualitatively new physical effects arise on these spatial scales because the PSOI proves to be attractive for a certain spin configuration of electrons tied to their momenta. The attraction has a clear origin. The PSOI due to the electric field of a given electron lowers down the energy of another electron, provided that the latter is in a particular spin orientation relative to its momentum. This effect gets stronger as the distance between electrons decreases, which means the attraction arises between the electrons. 

When the PSOI becomes larger than or comparable with the Coulomb repulsion, a wide scope opens up for many non-trivial effects both at the few-particle level and collective phenomena in the systems of many particles. First of all, it is obvious that the attractive interaction of two electrons can lead to the formation of a bound electron pair (BEP). There appears an interesting problem of their binding energy and possible types of spin-charge structure. This is a highly nontrivial issue, since the interaction is attractive only for certain spin and momentum configurations. Recent works in this direction~\cite{2018arXiv180410826G,PhysRevB.98.115137,10.1016/j.physe.2018.12.028,2019arXiv190409510G,bhzbeps} are discussed in Section~\ref{sec2}.

Section~\ref{sec3} is devoted to the problem of a many-particle correlated state formed owing to the PSOI\@. The presence of the Coulomb interaction simultaneously with the PSOI greatly complicates the problem, moreover, the studies carried out to date~\cite{PhysRevB.95.045138,GINDIKIN2022115328} show that the sufficiently strong PSOI leads to an instability of the homogeneous state of the system and therefore it is necessary to find out the adequate stabilizing mechanisms. At the present stage, studies were focused on the collective excitations under conditions where the PSOI is not too strong, but the system can approach the instability threshold. In this way, the spectrum and structure of the collective modes can be elucidated and the spectral functions of electronic fluctuations leading to the loss of stability can be found. Such studies were carried out for two specific situations of one-dimensional (1D) and two-dimensional (2D) systems.

In 1D quantum wires, collective excitations have been studied when the PSOI is produced by the image charges on a nearby gate~\cite{PhysRevB.95.045138}. It was found that the spectrum of collective modes contains two branches with a mixed spin-charge structure, the composition of which strongly depends on the magnitude of the PSOI\@. Of most interest is the behavior of one of them, which originates from the spinon branch in the absence of the PSOI\@. As the PSOI parameter increases, this mode strongly softens and ultimately leads to the instability of the system in the long-wavelength region of the spectrum. At the same time, its structure changes continuously from purely spin-like to purely plasmon-like.

In the direction of 2D systems, the situation of an electron gas with the mirror symmetry with respect to the plane was studied. In this case an unconventional correlated electronic state is found to arise due to the PSOI\@. This state is characterized by a sharp peak in the structure factor, indicating a tendency to form a striped structure with a certain spatial scale set by the competition between the Coulomb repulsion and the PSOI-induced attraction of electrons. The system becomes unstable on this scale if the density of electrons is larger than critical~\cite{GINDIKIN2022115328}. Interestingly, the fluctuations that grow rapidly as the system approaches the instability threshold consist mainly of the charge density fluctuations, just like in the 1D case.

In what follows, we review the PSOI manifestations in low-dimensional electron systems studied to date.

\section{Electron pairs bound by the spin-orbit interaction}
\label{sec2}

PSOI gives rise to a new mechanism of pairing of purely electronic nature that stems simply from the movement of electrons in certain spin configurations.

\begin{figure}
    \includegraphics[width=0.4\textwidth]{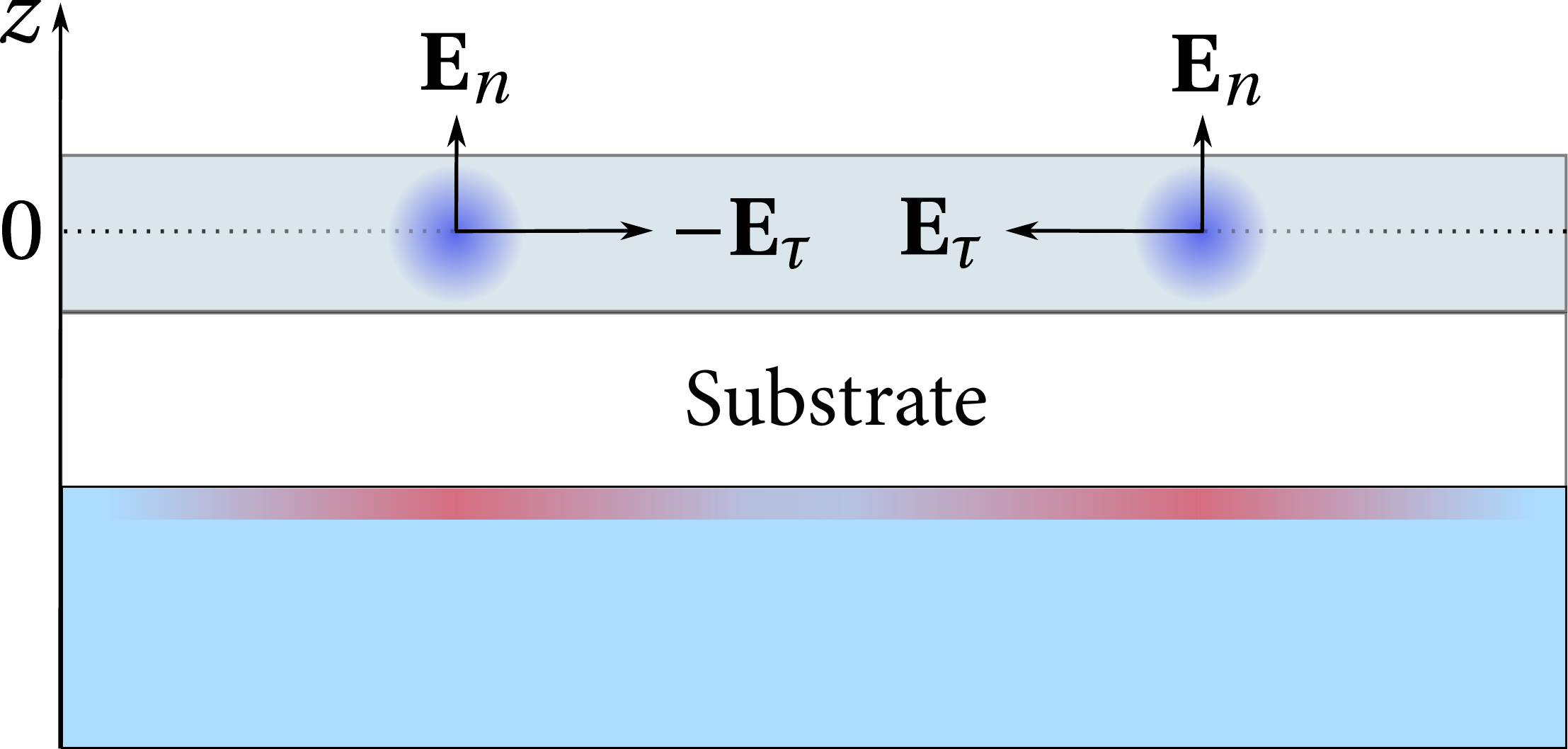} 
    \caption{A 2D layer with a proximate gate. Each electron experiences the electric fields from the neighboring electrons, the polarization charges, and the total charge of the gate.\label{pic_1}}
\end{figure}

A generic 2D electron system to study the BEPs appearing due to the PSOI can be viewed as a 2D layer proximated by a metallic gate, as shown in Fig.~\ref{pic_1}. 

The PSOI is created jointly by the in-plane $E_{\tau}(\bm{r})$ and normal $E_n(\bm{r})$ components of the electric field, where $\bm{r} \equiv (x,y,0)$ stands for the in-plane position. In the two-particle basis $\mathcal{B}=\{\lvert \uparrow \uparrow \rangle,\lvert \uparrow \downarrow \rangle,\lvert \downarrow \uparrow \rangle,\lvert \downarrow \downarrow \rangle\}$ the PSOI Hamiltonian is~\cite{2019arXiv190506340G}

\begin{widetext}
\begin{equation}
\label{ham2d}
    H_{\mathrm{PSOI}} =
        \frac{\alpha}{2 \hbar}
        \begin{pmatrix}
            \frac{4 E_{\tau}(\bm{r})}{r} {(\bm{r} \times \bm{p})}_{z} && -i\{E_{n}(\bm{r}),p_{-}\} + i E_{n}(\bm{r}) P_{-} && i\{E_{n}(\bm{r}),p_{-}\} + i E_{n}(\bm{r}) P_{-} && 0 \\
            i\{E_{n}(\bm{r}),p_{+}\} - i E_{n}(\bm{r}) P_{+} && \frac{2 E_{\tau}(\bm{r})}{r} {(\bm{r} \times \bm{P})}_{z} && 0 && i\{E_{n}(\bm{r}),p_{-}\} + i E_{n}(\bm{r}) P_{-} \\
            -i\{E_{n}(\bm{r}),p_{+}\} - i E_{n}(\bm{r}) P_{+} && 0 && -\frac{2 E_{\tau}(\bm{r})}{r} {(\bm{r} \times \bm{P})}_{z} && -i\{E_{n}(\bm{r}),p_{-}\} + i E_{n}(\bm{r}) P_{-} \\
            0 && -i\{E_{n}(\bm{r}),p_{+}\} - i E_{n}(\bm{r}) P_{+} && i\{E_{n}(\bm{r}),p_{+}\} - i E_{n}(\bm{r}) P_{+} && -\frac{4 E_{\tau}(\bm{r})}{r} {(\bm{r} \times \bm{p})}_{z}
        \end{pmatrix}\,,
\end{equation}
\end{widetext}
\noindent with $\bm{r} \equiv \bm{r}_1-\bm{r}_2$ being the relative position of two electrons, $\bm{R} \equiv (\bm{r}_1+\bm{r}_2)/2$ the barycenter position of the electron pair, $\bm{p} =\! -i\hbar \nabla_{\! \bm{r}}$ and $\bm{P} =\!-i\hbar \nabla_{\! \bm{R}}$ the corresponding momenta, $p_{\pm} = p_x \pm i p_y$. Curly brackets denote the anti-commutator.

Because of the translational invariance of the system the two-electron wave-function represents a product of two parts, one of which describes the free motion of the center of mass, and the other describes the relative motion of electrons within a pair, 
\begin{equation}
    \Psi(\bm{r}_1,\bm{r}_2) = e^{i \bm{K} \cdot \bm{R}} \psi_{\bm{K}}(\bm{r}) \equiv e^{i \bm{K} \cdot \bm{R}}{\left(\psi_{\uparrow \uparrow},\psi_{\uparrow \downarrow},\psi_{\downarrow \uparrow},\psi_{\downarrow \downarrow}\right)}^{\intercal}\,.
\end{equation}

It is important that the wave function of the relative motion $ \psi_{\bm{K}}(\bm{r})$, generally speaking, depends on the total momentum $K$ of the electron pair, since the binding potential of Eq.~\eqref{ham2d} depends on $K$.

Of special interest is a particular case of symmetric (non-gated) 2D systems, because it allows for an exact analytic solution. The symmetric 2D systems can be realized as freely suspended 2D layers, which are currently in the focus of the immense experimental activity aimed at enhancing the e-e interaction~\cite{Pogosov_2022}. In this case the equations of motion for the spinor components decouple, and we end up with two inherently different types of BEPs.

The motion of an electron pair as a whole generates a spin-orbit interaction, the magnitude of which is proportional to the momentum of the center of mass $\bm{K}$. This is given by the last term on left hand side of the equation of motion for the corresponding spinor components $\psi_{\uparrow \downarrow}$ and $\psi_{\downarrow \uparrow}$ 
\begin{equation}
\label{conv_2d}
\begin{split}
        &\left[- \frac{\hbar^2}{m} \nabla^2_{\bm{r}} - \frac{\hbar^2}{4m} \nabla^2_{\bm{R}} + U(\bm{r}) + \frac{\alpha}{\hbar} \frac{E_{\tau}(\bm{r})}{r} {(\bm{r} \times \bm {P})}_z \right] \psi_{\uparrow \downarrow} \\
        &= \varepsilon_{\uparrow \downarrow} \psi_{\uparrow \downarrow}\,.
\end{split}
\end{equation}
Equation for $\psi_{\downarrow \uparrow}$ is similar with a sign change before $\alpha$. The total wave function, antisymmetric with respect to electron interchange, represents a mixed singlet-triplet state. The binding potential is anisotropic, since the rotational symmetry in the plane is broken by the presence of a preferred direction along $\bm{K}$. Hence the wave function acquires a non-trivial angular dependence, which for the case a purely Coulomb in-plane field is of the form
\begin{equation}
\label{convspinor}
\begin{split}
   \Psi(\bm{r},\bm{R}) ={}& e^{i \bm{K} \cdot \bm{R}} \bigg( 0, ce_0 (\tfrac{\phi}{2},2 \tilde{\alpha} K a_B ) g(r),\\
     {}&-ce_0 (\tfrac{\phi + \pi}{2},2 \tilde{\alpha} K a_B ) g(r), 0{\bigg)}^{\intercal} \,,
\end{split}
\end{equation}
where $\phi$ is the polar angle measured from the $\bm{K}$ direction, $ce_{0}(z,q)$ is the Mathieu function~\cite{olver}, the radial part $g(r)$ is given in Ref.~\cite{PhysRevB.98.115137}, and the dimensionless Rashba constant $\tilde{\alpha} = \alpha/e a_B^2$ is normalized on the Bohr radius $a_B$ in the material. 

\begin{figure}
        \includegraphics[width=0.5\textwidth]{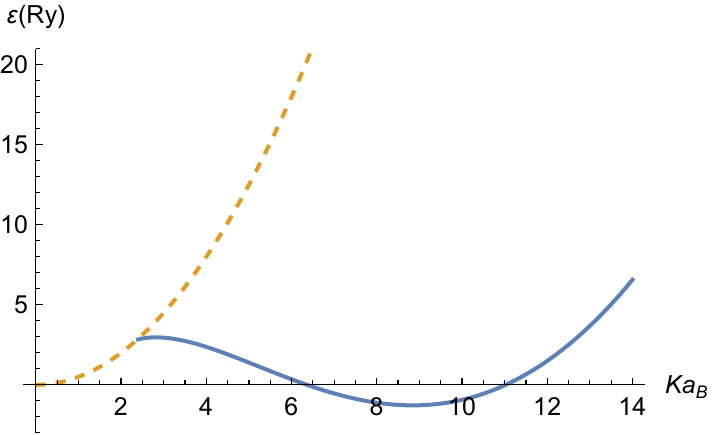}
        \caption{The energy level of the convective BEP (solid line) and the kinetic energy of the center of mass (dashed line) vs $K a_B$ for $\tilde{\alpha} =1 $.\label{pic_2}}
\end{figure}

BEPs of this type arise as soon as the center-of-mass momentum $K$ exceeds a critical value. For this reason this solution is called convective. Its spectrum is illustrated in Fig.~\ref{pic_2}. The binding energy grows with $K$ so that the total energy of the pair starts to decrease with $K$, which can even lead to the negative effective mass of the BEP in a certain region of $K$.

On the other hand, two electrons can perform the relative motion, rotating around a common barycenter. This motion also gives rise to a PSOI, as described by the last term on the left hand side of the equation of motion for the corresponding spinor components,
\begin{equation}
\label{rel_2d}
\begin{split}
    	&\left [- \frac{\hbar^2}{m} \nabla^2_{\bm{r}} - \frac{\hbar^2}{4m} \nabla^2_{\bm{R}} + U(\bm{r}) + \frac{2\alpha}{\hbar} \frac{E_{\tau}(\bm{r})}{r} {(\bm{r} \times \bm {p})}_z \right ] \psi_{\uparrow \uparrow}\\
    	&= \varepsilon_{\uparrow \uparrow} \psi_{\uparrow \uparrow}\,,
\end{split}
\end{equation}
and similarly for $\psi_{\downarrow \downarrow}$, but with with a sign change before $\alpha$. This gives rise to the relative BEPs, which appear in degenerate pairs of triplet states formed by electrons with parallel spins locked to the angular momentum direction,
\begin{equation}
\label{deg1}
    \psi_{+}(\bm{r}) = {\left(u(r) e^{- i \phi},0,0,0\right)}^{\intercal}
\end{equation}
and 
\begin{equation}
\label{deg2}
    \psi_{-}(\bm{r}) = {\left(0,0,0,u(r) e^{i \phi}\right)}^{\intercal}\,.
\end{equation}
In contrast to the convective BEPs, the barycenter motion is fully decoupled from the relative motion of electrons and hence has no impact on the structure and spectrum of the relative BEPs.

\begin{figure}
        \includegraphics[width=0.45\textwidth]{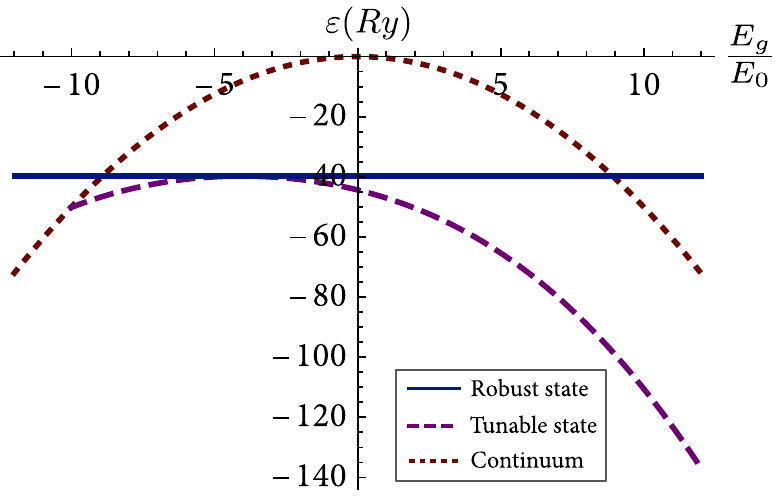} 
        \caption{The binding energy of the robust and tunable BEPs as well as the continuum boundary vs.\ the gate field $E_g$ normalized at $E_0 = e/2\epsilon a_B^2$.\label{pic_3}}
\end{figure}

The presence of the gate lifts the twofold spin degeneracy of the relative BEPs. The interplay of the in-plane $E_{\tau}(r)$ and normal $E_n(r)$ electric fields leads to a profound rearrangement of the bound states. In the presence of both field components the relative motion of electrons and the motion of their center of mass are strongly coupled. The BEPs formation and structure were investigated for a particular case of the electron pairs with zero total momentum~\cite{2019arXiv190409510G}. The degenerate ground state splits into a pair of states with rather different properties. 

There is a tunable state
\begin{equation}
    \label{BS}
        \psi(\bm{r}) = {\left(v(r)e^{- i \phi}, w(r), -w(r), v(r) e^{i \phi}\right)}^{\intercal}
\end{equation}
whose binding energy grows with a gate voltage, with its orbital and spin structure continuously changing. At large negative gate voltage the tunable state gets into the continuum of states and delocalizes. 

There also appears a robust state
\begin{equation}
    \label{prot}
            \psi(\bm{r}) = {\left(u(r)e^{- i \phi},0,0, -u(r) e^{i \phi}\right)}^{\intercal}\,,
\end{equation}
which does not change its orbital and spin structure upon the change of the gate voltage. Its energy level eventually gets in the conduction band for sufficiently high gate voltage of any sign, where the robust state remains bound and localized. In other words, a bound state in continuum appears with a corresponding $\delta$-peak in the density of states. Fig.~\ref{pic_3} shows the energy levels of the robust state and tunable bound state plotted against the normalized field of the gate~\cite{2019arXiv190409510G}.

Similar physical picture holds for the BEPs formed in gated 1D quantum wires~\cite{2018arXiv180410826G}. The binding energy of the BEPs lie in the meVs range for materials with giant Rashba effect. Most interestingly, they are tunable by the gate voltage. However, huge values of the Rashba constant of the material of the order of $\tilde{\alpha} \approx 1$ are necessary in order for the BEPs to appear.

\begin{figure}[htb]
    \includegraphics[width=1.0\linewidth]{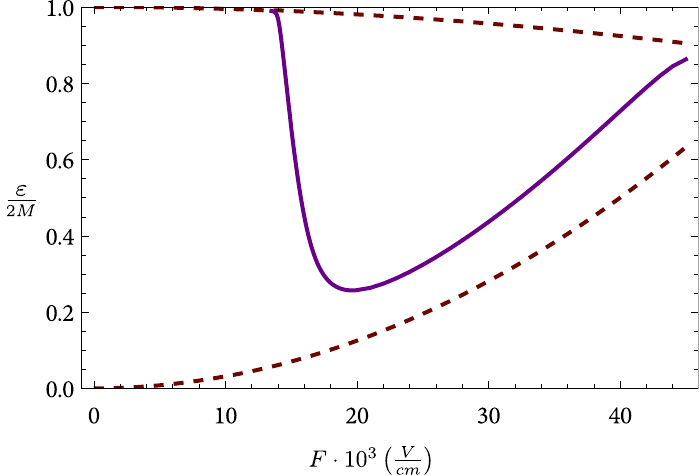}
    \caption{The BEP energy level as a function of the gate electric field $F$. The continuum boundaries are shown by the dashed lines.\label{pic_4}}
\end{figure} 

We emphasize that the SOI results from the hybridization of basic Bloch states with different spin configurations. For this reason the BEPs formed in a system with extremely strong SOI should be, generally speaking, studied in the frame of the multi-band model that properly takes into account the hybridization of various electron orbitals forming the BEP\@.

To demonstrate the formation of BEPs in multi-band systems, two-body problem was solved within the frame of the four-band model of Bernevig, Hughes and Zhang~\cite{Bernevig1757} in the  presence of the Rashba SOI~\cite{Rothe_2010}. In this case, the BEPs arising due to the PSOI are characterized only by the total angular momentum along the normal direction, while the spin is not defined~\cite{bhzbeps}. Their energy depends on the gate electric field $F$ producing the Rashba SOI\@. Figure~\ref{pic_4} shows the ground-state energy of a BEP with the total angular momentum $l=0$ as a function of $F$. The energy level lies in the gap of the two-particle spectrum which also depends on $F$. Note the non-monotonous dependence of the BEP energy on the gate field, which strongly affects the hybridization of the electron bands that form the BEP\@.

\section{Correlated electron state formed owing to the PSOI}
\label{sec3}

Of greatest interest are the effects due to the PSOI arising in many-electron systems, which researchers deal with in currently known low-dimensional materials with a strong Rashba SOI\@. The presence of a large number of electrons significantly affects the PSOI interaction, since electron correlations play a fundamental role in the formation of the electric field that determines the PSOI\@.

To be specific, we start with a 2D electron system with the mirror symmetry with respect to the plane, where the PSOI is generated by the in-plane Coulomb field. In this case the PSOI Hamiltonian of Eq.~\eqref{hampsoi} takes the following form,
\begin{align}
   \hat{H}_{\mathrm{PSOI}} = \frac{2 m \alpha}{e \hbar} \int d\bm{r}_1 d\bm{r}_2 :\mspace{-3mu} \hat{\rho}(\bm{r}_1) {\left( \bm{\mathcal{E}}(\bm{r}_1 - \bm{r}_2) \times  \hat{\bm{j}}_{\sigma}(\bm{r}_2)\right)}_z \mspace{-3mu}:\,,
\end{align}
where $::$ denotes the normal ordering, $\hat{\rho}(\bm{r})$ is the density operator, and $\hat{\bm{j}}_{\sigma}(\bm{r})$ is the paramagnetic part of the spin current,
\begin{align}
    \hat{\bm{j}}_{\sigma}(\bm{r}) = \frac{i e \hbar}{2m} \sum_{\varsigma} \varsigma \left[\hat{\psi}^{+}_{\varsigma}(\bm{r}) \nabla_{\!\bm{r}} \hat{\psi}_{\varsigma}(\bm{r}) - (\nabla_{\!\bm{r}} \hat{\psi}^{+}_{\varsigma}(\bm{r})) \hat{\psi}_{\varsigma}(\bm{r})\right]\,.
\end{align}

With the PSOI present, the effective e-e interaction strength, defined by the ratio of the interaction energy to the Fermi energy, is characterized by two parameters instead of one. While for the Coulomb interaction the interaction strength is characterized by the parameter $r_s$, which is the ratio of the inter-electron distance to the Bohr radius $a_B$, the PSOI is described by another parameter, $\tilde{\alpha}/r_s$.

\begin{center}
    \begin{tabular}{ccc}
        Interaction parameter & $H_{\mathrm{Coul}}$ & $H_{\mathrm{PSOI}}$\\
        \midrule
          $\dfrac{E_{\mathrm{int}}}{E_{\mathrm{kin}}}$ & $r_s$ & $\dfrac{\tilde{\alpha}}{r_s}$
    \end{tabular}
\end{center}

We emphasize that these two interaction parameters exhibit different dependence on the parameters of the electronic system. Thus, the PSOI parameter $\tilde{\alpha}/r_s$ increases with the electron density, in contrast to the Coulomb interaction parameter $r_s$. The PSOI parameter exceeds $r_s$ at a sufficiently high density, $n>1/(\tilde{\alpha}a_B^2)$. In materials with a weak SOI, the critical density $n_c=1/(\tilde{\alpha}a_B^2)$ is unrealistically high, but in materials with a strong SOI, the PSOI becomes significant and even dominant at a quite acceptable electron density. Of course, this density cannot exceed a maximum value, which depends on the specific structure of the material, but can be roughly estimated at the level of $\SI{e14}{cm^{-2}}$. 

The charge susceptibility calculated within the random phase approximation (RPA) is equal to~\cite{GINDIKIN2022115328}
\begin{equation}
    \label{chinn}
        \chi_{nn}(q,\omega) =\frac{\chi_0(q,\omega)}{1 - \mathcal{U}_q [1 - \mathcal{G}_{\mathrm{PSOI}}(q,\omega)] \chi_0(q,\omega)}\,.
\end{equation}
Here $\chi_{0}(q,\omega)$ stands for usual 2D Lindhard susceptibility~\cite{PhysRevLett.18.546}. The effect of the PSOI is described by the dynamic local field correction
\begin{equation}
        \mathcal{G}_{\mathrm{PSOI}}(q,\omega) = - \frac{8 \alpha^2}{e^2} \mathcal{U}_q\, \chi_{j_{T}j_{T}}(q,\omega)\,,
\end{equation}
which is related to the transverse current-current susceptibility
\begin{align}
    \chi_{j_{T}j_{T}}(q,\omega) ={}& N_{\sigma} k_F^4 \frac{q}{k_F} \times\\
        & \Big( \frac{\nu_{-}}{2} - \frac13 \mathrm{sign} (\mathrm{Re} \nu_{-}) \left[{(\nu_{-}^2)}^{\frac32} - {(\nu_{-}^2-1)}^{\frac32}\right]\notag\\
        {}&- \frac{\nu_{+}}{2} + \frac13 \mathrm{sign} (\mathrm{Re} \nu_{+}) \left[{(\nu_{+}^2)}^{\frac32} - {(\nu_{+}^2-1)}^{\frac32}\right] \Big)\,,\notag
\end{align}
with $ \nu_{\pm} = \frac{\omega +i0}{q v_F} \pm \frac{q}{2k_F}$, and $N_{\sigma} = \frac{m}{2 \pi \hbar^2}$ being the density of states at the Fermi-surface. 

The static structure factor $S(q)$, related to the susceptibility of Eq.~\eqref{chinn}, has a singularity at a critical value of the $r_s=r_s^*$,
\begin{equation}
    r_s^* = \frac{2^{\frac{13}{6}} \tilde{\alpha}}{\sqrt{2^{\frac13} +2 {\tilde{\alpha}}^{\frac23}3^{\frac23}}}\,.
\end{equation}
As we approach the critical value $r_s \to r_s^* +0$, the spectral weight is being transferred to the long-wavelength part of the spectrum, with the sharp peak formed in $S(q)$ at the critical value $q_c$ of the wave-vector, equal to
\begin{equation}
    \frac{q_c}{k_F} = \frac{2 {\tilde{\alpha}}^{\frac13} 3^{\frac13}}{\sqrt{2^{\frac13} +2 {\tilde{\alpha}}^{\frac23}3^{\frac23}}}\,.
\end{equation} 

\begin{figure}[htp]
	\includegraphics[width=1.0\linewidth]{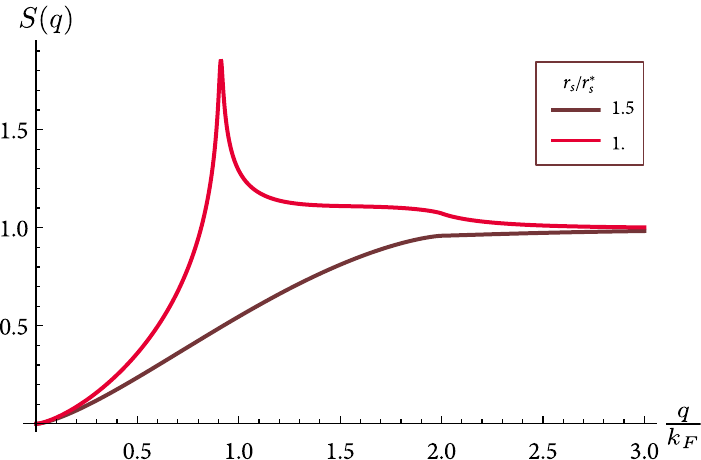}
	\caption{The structure factor $S(q)$ as a function of $q$ for two values of the $r_s$ parameter. The Rashba constant is $\tilde{\alpha}=0.1$, which corresponds to $r_s^*=0.3$.\label{pic_5}}
\end{figure}

This behavior, illustrated by Fig.~\ref{pic_5}, indicates the appearance of strong electron correlations due to the PSOI on a new spatial scale $l=q_c^{-1}$ set by the competition between the PSOI-induced attraction of electrons and their Coulomb repulsion. For typical values of the SOI parameter $\tilde{\alpha} \ll 1$ found in Rashba materials, the critical value $q_c \propto \tilde{\alpha}^{1/3} k_F$ lies in the long-wave part of the spectrum, whereas $q_c \to \sqrt{2} k_F$ when the PSOI is extremely strong. 

\begin{figure}[htp]
	\includegraphics[width=0.9\linewidth]{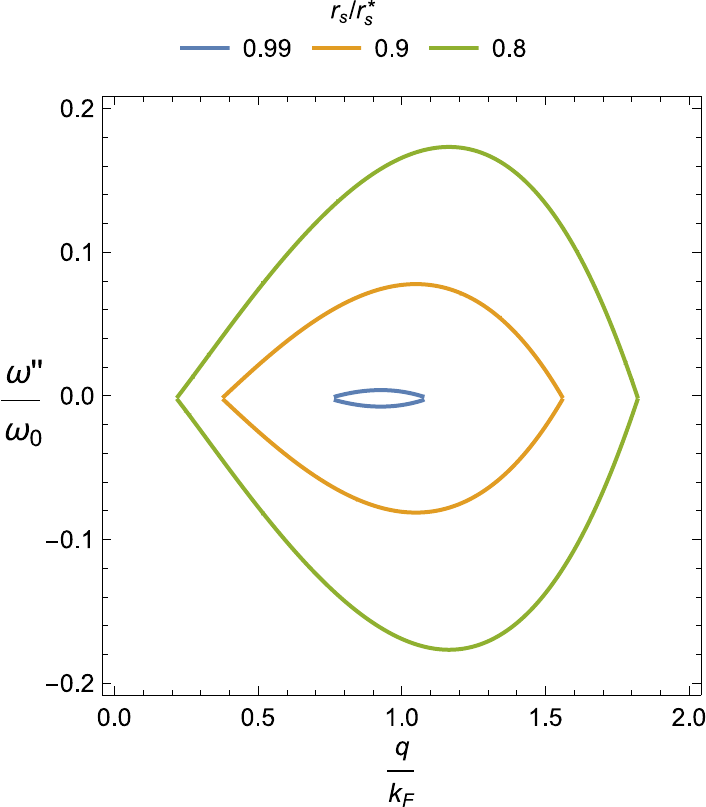}
	\caption{The imaginary part of the frequency of a new solution of the dispersion equation due to the PSOI as a function of wave vector. The dispersion line is shown for three values of $r_s$ to trace how the instability develops in the system with increasing the PSOI interaction parameter of $\tilde{\alpha}/r_s$. The Rashba constant equals $\tilde{\alpha}=0.1$, which corresponds to $r_s^*=0.3$.\label{pic_6}}
\end{figure}

It is interesting that at $r_s \le r_s^*$ a new solution of the dispersion equation for the collective modes arises due to the PSOI\@. The solution exists in a finite band of wave vectors around $q_c$, the width of which grows with $\tilde{\alpha}/r_s$. The spectrum of this branch of solution is plotted in Fig.~\ref{pic_6}.

The frequency of the solution is purely imaginary, which means that the electron density fluctuations are exponentially growing with time. In other words, the spatially uniform paramagnetic state of the electron system becomes unstable with respect to the charge density fluctuations. This indicates a tendency for the formation of a striped structure on the $q_c$ scale. A specific type of the electron state corresponding to the true energy minimum should be found within more advanced approaches that allow one to adequately take into account the nonlinear processes. 

The other system where the PSOI is created by the normal component of the electric field of the e-e interaction demonstrates a different scenario of the instability, which also appears at a large parameter of the Rashba SOI\@. The system is a 1D quantum wire with a nearby gate on which the interacting electrons induce the image charges. In this case the normal electric field of the image charges is the only source of the PSOI, which strongly affects the collective excitations.

In a usual 1D system with the Coulomb e-e interaction the collective excitations are known to be plasmons and spinons, which are separated according to a fundamental property of the Tomonaga-Luttinger liquid~\cite{voit1995one}. The plasmon velocity is renormalized by the Coulomb repulsion, which increases the system stiffness, whereas the speed of spinons is practically independent of the Coulomb interaction.

The theory of the collective excitations  with the PSOI generated by the image charges was developed using two approaches: the  bosonization~\cite{PhysRevB.95.045138,2017arXiv170700316G} and RPA~\cite{PhysRevB.95.045138,doi:10.1002/pssr.201700256}, which led to close results.

The collective excitation spectrum has two branches, the frequencies $\omega_{\pm}$ of which are
\begin{align}
\label{dispersion}
    {\left(\frac{\omega_{\pm}}{q v_F}\right)}^2 ={}& 1 + \left( \mathcal{U}_q -\alpha_*^2 \mathcal{F} \mathcal{E}_q \right)\\
    {}& \pm \sqrt{ {\left(\mathcal{U}_q - \alpha_*^2 \mathcal{F} \mathcal{E}_q \right)}^2 + \alpha_*^2 \mathcal{E}_q^2}\,,\notag
\end{align}
with the Fermi velocity $v_F = \hbar k_F/m$, the PSOI interaction parameter $\alpha_* = \tilde{\alpha}/(\pi r_s)$, the Fourier-transformed Coulomb interaction $U(q) = 2(e^2/ \epsilon) \left(K_0(qd) - K_0(qa)\right)$ and normal field $E(q) = 2 (e/\epsilon) |q| K_1(|q|a)$ normalized according to $\mathcal{U}_q = U(q)/(\pi \hbar v_F)$ and $\mathcal{E}_q = \epsilon E(q)/(e n_0)$; $d$ is the quantum wire diameter, $a/2$ is the separation between the wire and the gate, $n_0$ is the mean electron density. The electric field $F = E(0) + 4 \pi n_g/\epsilon$ produced by electron's own image and the total charge $n_g$ of the gate is normalized as $\mathcal{F} = F/(e n_0^2)$.

\begin{figure}
    \centering
    \includegraphics[width=0.5\textwidth]{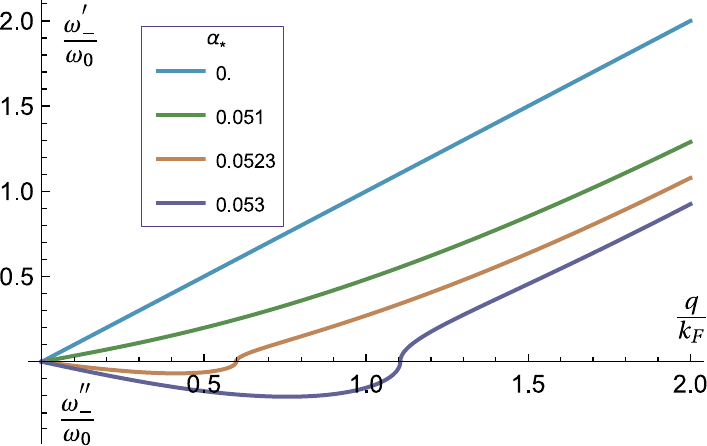} 
    \caption{The real ($\omega'_{-}$) and imaginary ($\omega''_{-}$) parts of the frequency of the collective mode as a function of the wave vector for several values of the PSOI constant. The frequency is normalized to $\omega_0 = v_F k_F$.\label{pic_7}}
\end{figure}

The main effect that the PSOI produces is with the mode $\omega_{-}$. Fig.~\ref{pic_7} allows one to trace its behavior as $\alpha_*$ grows. The mode velocity goes down to zero at the critical value of PSOI amplitude, which equals
\begin{equation}
    \label{alphat}
        \alpha_c(q) = \dfrac{\sqrt{1 + 2\mathcal{U}_q}}{\sqrt{\mathcal{E}_q^2 + 2\mathcal{\mathcal{F}}\mathcal{E}_q}}\,.
\end{equation}

For $\alpha_* > \alpha_c$ the mode frequency becomes imaginary, which is a sign of the system instability with respect to the long-wave density fluctuations. Again, the instability develops in the charge sector of the system, which can be seen from the charge stiffness 
\begin{equation}
    \label{stiffness}
        \varkappa = \pi\hbar v_F (1 + 2\mathcal{U}_{0}) \left( 1 - \frac{\alpha_*^2}{\alpha_c^2(0)} \right)
\end{equation}
going to zero at $\alpha_* = \alpha_c(0)$.

Another effect of the PSOI in 1D electron systems is that it violates the spin-charge separation. The spin and charge degrees of freedom of the collective excitations become intertwined~\cite{PhysRevB.95.045138,doi:10.1002/pssr.201700256,2017arXiv170700316G}. The spin-charge composition of the collective mode can be characterized by a spin-charge separation parameter $\xi$, essentially defined as the degree of the spin polarization in the mode
\begin{equation}
	\xi_{\pm} \equiv \frac{n_{\uparrow} + n_{\downarrow}}{n_{\uparrow} - n_{\downarrow}} = \frac{v_{\pm} - v_{\pm}^{-1}}{\alpha_* \mathcal{E}_q} \,,
\end{equation}
where we introduced the mode phase velocity $v_{\pm} = \omega_{\pm}/q v_F$.
 
\begin{figure}
    \includegraphics[width=0.5\textwidth]{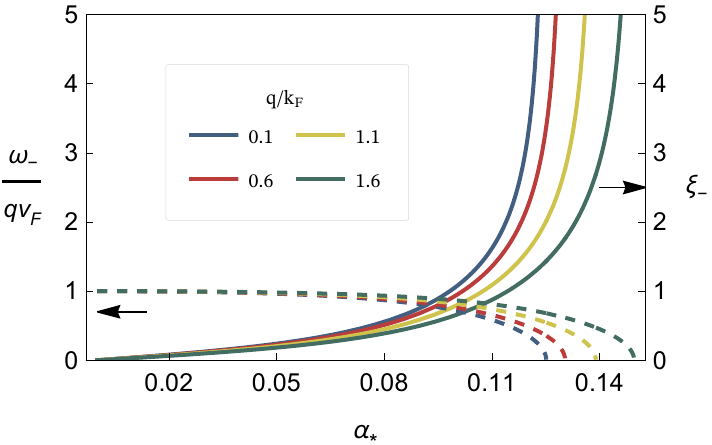}
    \caption{The spin-charge separation parameter (solid line) and normalized phase velocity (dashed line) for the $\omega_{-}$  branch of collective excitations as a function of the PSOI interaction parameter for several $q$.
    \label{pic_8}}
\end{figure}

At $\alpha_* = 0$ we have $\xi_{-} = 0$, that is $\omega_{-}$ corresponds to a purely spinon excitation ($n_{\uparrow} = - n_{\downarrow}$) with the energy dispersion $\omega_{-} = v_F q$ not affected by the interactions. The mode velocity $v_{-}(q) \to 0$ turns to zero in the long-wavelength region of the spectrum as $\alpha_* \to \alpha_c(q)$, whereas the spin-charge separation parameter $\xi_{-} \to \infty$ diverges. In other words, the collective mode $\omega_{-}$ turns into a purely plasmon excitation ($n_{\uparrow} = n_{\downarrow}$) at the instability threshold $\alpha_* = \alpha_c(q)$. The evolution of the spin-charge structure  of the mode with the change in $\alpha_*$ is traced in Fig.~\ref{pic_8}.  

The presence of the PSOI makes it possible to control both $v_{\pm}$ and $\xi_{\pm}$ via the gate electric field $\mathcal{F}$ by tuning the gate potential. In particular, $v_- \to 0$ as $\mathcal{F}$ grows, that is the corresponding mode softens. Without PSOI, the gate voltage has no effect on the excitation velocities nor does it violate the spin-charge separation between the modes, because the one-body Rashba SOI produced by the gate field can be completely eliminated in 1D by a unitary transformation~\cite{PhysRevB.88.125143}.

The violation of the spin-orbit separation results in electron transport features that can be observed to reveal the PSOI effects from purely electrical measurements. Since both spin-charge-mixed modes of a system with broken spin-charge separation convey the electric charge, both contribute to the electric response of the system. Thus, the dynamic admittance of the 1D quantum wire coupled to leads has features due to presence of the both types of the collective excitation in contrast to the system with purely Coulomb interaction.

The quantum wire admittance was found to be~\cite{2017arXiv170700316G}
\begin{equation}
\label{admit}
    \frac{G_{\omega}}{G_0} = \frac{1 - v_{-}^2}{v_{+}^2 - v_{-}^2} \frac{1}{1 - i v_{-} \tan \frac{\omega \tau}{2 v_{-}}} + \frac{v_{+}^2 - 1}{v_{+}^2 - v_{-}^2} \frac{1}{1 - i v_{+} \tan \frac{\omega \tau}{2 v_{+}}}\,.
\end{equation}
Here the conductance quantum is $G_0 = 2e^2/h$, the characteristic transit time is $\tau = L/v_F$, with $L$ being the wire length. 

\begin{figure}
    \includegraphics[width=0.5\textwidth]{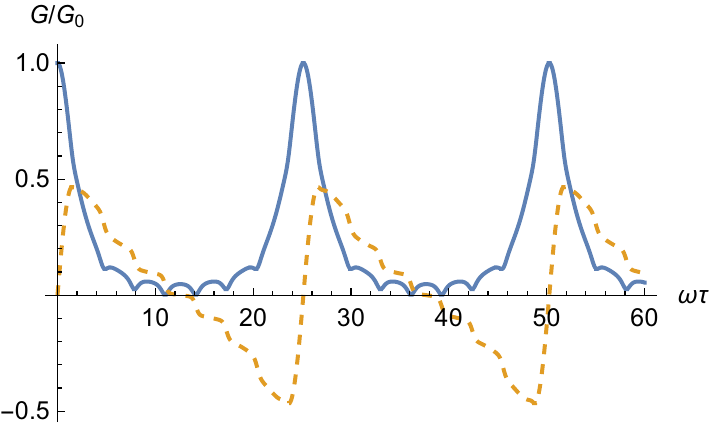} 
    \caption{The real (solid line) and imaginary (dashed line) admittance components vs.\ frequency.
    \label{pic_9}}
\end{figure}

Two terms on the right hand side of the Eq.~\eqref{admit} reflect the contributions of each of the collective modes. The frequency dependence of $G_{\omega}$ plotted  in Fig.~\ref{pic_9} features a specific double-periodic oscillatory pattern, corresponding to the Fabry-Pérot resonances at the wire length. Two distinct characteristic frequencies result from different transit times of the slow and fast collective modes. Both mode velocities, renormalized by the PSOI, could be extracted from the measurements of the characteristic oscillations of the admittance using the methods of Ref.~\cite{Chudow2016}.

\section{Conclusion and outlook}
The pair spin-orbit interaction between the electrons, which is usually considered relativistically weak, becomes extremely strong in modern materials with a strong Rashba spin-orbit interaction. The giant PSOI in Rashba materials is of slightly different nature as compared to that known in relativistic quantum mechanics, although it certainly has a relativistic origin. It arises because of the Rashba effect — the combined effect of the momentum-dependent hybridization of the electron subband with other subbands split due to the intra-atomic spin-orbit interaction, and the inversion symmetry breaking produced by an electrical field, in this case, the Coulomb field of interacting electrons.

The PSOI has two main features:
\begin{itemize}
    \item
     The PSOI leads to the attraction of electrons for certain configurations of the spins relative to their momenta.
    \item
     The magnitude of the PSOI is determined by the electric field of the interacting electrons and hence increases with decreasing distance between them much faster than the interaction potential.
\end{itemize}

Owing to these features, strong PSOI opens up wide prospects for the appearance of many nontrivial effects due to the formation of new strongly correlated states of electrons, which are still very poorly understood. According to the studies conducted to date, such effects are manifested for sufficiently strong Rashba SOI, which is apparently attained in the already available materials. Because of the impressive progress in the creation of new low-dimensional materials with a giant SOI, theoretical studies of such states are very important.

Nontrivial effects arise at the level of systems with few particles, as well as in many-particle systems.

PSOI leads to the formation of bound pairs of electrons with various properties, depending on the screening conditions and the presence of the SOI induced by the gate potential. An attractive feature of such states is the possibility to achieve a high binding energy of the pairs and to control their spectrum. Prospects for further research in this direction are associated primarily with the investigation of the stability of such pairs with respect to the interaction with other electrons and to the formation of more complex electron aggregations.

In many-particle systems the PSOI, even if not very strong, dramatically affects the spectra of the collective excitations and their spin-charge structure. Of particular interest is the softening of the collective modes due to the attraction between the electrons in certain spin configurations, which is produced by the PSOI and is accompanied by a rearrangement of the structure of electron correlations. Such studies for 2D systems at the current stage reveal a lot of interesting things. But of most importance is the fact that at a sufficiently strong PSOI the uniform state of a many-electron system becomes unstable. The structure of growing fluctuations indicates a tendency towards the formation of an inhomogeneous stripe phase. The search for the stabilization mechanisms and possible structures of the stable state is an open question most promising for further research.

\begin{acknowledgments}
Y.G.\ gratefully acknowledges the support and hospitality of the Weizmann Institute of Science during his stay there. The work of V.A.S.\ was carried out in the framework of the state task for the Kotelnikov Institute of Radio Engineering and Electronics
and partially supported by the Russian Foundation for Basic Research, Project No. 20--02--00126.
\end{acknowledgments}

\bibliography{paper}

\end{document}